\documentclass[preprint]{revtex4}
\usepackage[T1]{fontenc}
\usepackage{graphicx}
\usepackage{rotating}
\usepackage{bm}        
\usepackage{amssymb}   
\usepackage{bm}
\usepackage{float}
\usepackage{tikz}
\usepackage{diagbox}
\usepackage{amsmath}

\def\k1{k_1}
\def\k2{k_2}
\def\q1{q_1}
\def\q2{q_2}

\def\({\left (}
\def\){\right )}
\def\[{\left [}
\def\]{\right ]}

\newcommand{\beq}{\begin{equation}}
\newcommand{\eeq}{\end{equation}}
\newcommand{\bra}[1]{\langle #1 |}
\newcommand{\ket}[1]{| #1 \rangle}

\begin{document}
\date{\today}
\flushbottom \draft
\title{ 
Quantum transfer through small networks coupled to phonons: effects of topology vs phonons
}
\author{B. Hou$^{a}$ and R. V. Krems$^{a,b}$}
\affiliation{
$^a$Department of Chemistry, University of British Columbia, Vancouver, B.C. V6T 1Z1, Canada \\
$^b$Stewart Blusson Quantum Matter Institute, University of British Columbia, Vancouver, British Columbia, Canada V6T 1Z4
}

\begin{abstract}

Particle or energy transfer through quantum networks is determined by network topology and couplings to environments. 
This study examines the combined effect of topology and external couplings on the efficiency of directional quantum transfer through quantum networks.  
We consider a microscopic model of qubit networks coupled to external vibrations by Holstein and Peierls couplings. 
By treating the positions of the network sites and the site-dependent phonon frequencies as independent variables, we determine the Hamiltonian parameters corresponding to minimum transfer time by Bayesian optimization. 
The results show that Holstein couplings may accelerate transfer through sub-optimal network configurations but cannot accelerate quantum dynamics beyond the limit of the transfer time in an optimal phonon-free configuration. 
By contrast, Peierls couplings distort the optimal networks to accelerate quantum transfer through configurations with less than six sites. 
However, the speed-up offered by Peierls couplings decreases with the network size and disappears for networks with more than seven sites. For networks with seven sites or more, Peierls couplings distort the optimal network configurations and change the mechanism of quantum transfer but do not affect the lower limit of the transfer time. The machine-learning approach demonstrated here can be applied to determine quantum speed limits in other applications. 

\end{abstract}

\maketitle

\newpage

\section{Introduction}
Understanding dynamics of quantum transfer through disordered, finite-size networks is important for understanding  spin and charge transport in quantum mesoscopic materials \cite{spin-charge1,spin-charge2,spin-charge3}, electron and exciton transfer through molecular wires \cite{wires1, wires2}, transfer of excitons to reaction centres in photosynthetic complexes \cite{photosyn1,photosyn2}, dynamics of quantum spins on graphs \cite{spingraph}, quantum random walks on graphs \cite{qwalk1,qwalk2,qwalk3} and the design of quantum circuits for quantum computation \cite{qcircuit1,qcircuit2,qcircuit3}. Previous studies of quantum dynamics in small networks have considered the role of network topology \cite{complex-networks-from-classical-to-quantum,network1,network2} and the role of an external environment \cite{qenvironment1, qenvironment2,qenvironment3}. 
In particular, the experimental measurements of long-lived coherences in biological complexes stimulated much recent work exploring the possibility of environment-induced coherences and environment-assisted quantum transport \cite{EAQT1, EAQT2, EAQT3, EAQT4}. While environments generally lead to decoherence, it has been demonstrated that specific non-Markovian baths can accelerate quantum transfer through seven-member networks of Fenna-Matthew-Olson (FMO) photosynthetic aggregates \cite{Non-Mar1, Non-Mar2,Non-Mar3, EAQT3}. Similar observations have been made with qubit networks based on trapped ions subjected to controlled non-Markovian noise \cite{EAQT1}. The work with FMO aggregates has led to speculations that natural evolution has created the environments in photosynthetic complexes to be conducive to quantum transport of Frenkel excitons \cite{qevolution}. However, the combined effects or competition between topology and couplings to external systems have, to the best of our knowledge, not been investigated. 

The present work is stimulated by the following question: if nature could evolve both the topology and the environment of a quantum network, what would it choose to accelerate quantum transport? 
In addition to implications for energy transfer in photosynthetic aggregates, this question is important for the design of microscopic quantum devices for various applications \cite{qdevice1,qdevice2,qdevice3,qdevice4}.
To address this question, it is necessary to consider quantum dynamics in networks with variable topology coupled to variable environments. 
This problem is difficult because the number of possible network configurations and external degrees of freedom grow quickly with the system/environment size, requiring an increasing number of quantum dynamics calculations with increasing difficulty to explore the solutions of the Schr\"{o}dinger equation in the entire Hamiltonian parameter space. However, the evolution of quantum solutions in high-dimensional parameter spaces can be efficiently explored with machine learning, as was recently demonstrated for several problems \cite{mldim1,mldim2,mldim3,mldim4,mldim5,mldim6}.  

Here, we build a microscopic model, where every degree of freedom, including  the positions of network sites and the couplings of each site to an external phonon bath, is a variable in an optimization problem. 
We combine rigorous quantum dynamics calculations with a probabilistic machine learning approach that is designed to optimize simultaneously the  network configurations and the individual site couplings to the external environment. 
More specifically, we consider three-dimensional networks of $N=4-8$ qubits, with each qubit coupled either to Holstein phonons  \cite{Holstein1, Holstein2} or Peierls phonons  \cite{Peierls}. An excitation is injected into the network through an input site and collected by a non-Hermitian boundary at the output site. 
By treating the three-dimensional positions of qubits and effective phonon couplings as independent variables, we formulate a high-dimensional optimization problem that minimizes the excitation transfer time. The minimization is achieved by building Gaussian process (GP) models \cite{rasmussen-book} of the dependence of the transfer time on the independent variables and exploiting the properties of the resulting GPs for Bayesian optimization (BO) \cite{BO}. 

This work builds on a series of papers that examined the role of the spatial arrangement of FMO sites on exciton transport through photosynthetic aggregates \cite{optnet1,optnet2,efftrans,2d-Peierls}. 
The work of Scholak \textit{et al.} \cite{optnet1} has also examined in detail the effect of the non-Hermitian sink on the excitation transfer dynamics. We use the work of Scholak \textit{et al.} \cite{optnet1} to calibrate our results, illustrate the performance of BO and choose the optimal sink Hamiltonian parameters. The present study also builds on the work by Kurt \textit{et al.} \cite{efftrans} that examined the effect of Holstein phonons and the work of Mozafari \textit{et al.} \cite{2d-Peierls} that examined the effect of Peierls phonons on energy transfer through small networks. Our intermediate results for phonon-free networks or networks with fixed site positions are consistent with previous observations \cite{optnet1}. 
Our global optimization results illustrate that (1) sub-optimally arranged networks can benefit from both Holstein and Peierls phonons; (2) optimal networks benefit from Peierls phonons for $N \leq 5$; (3) optimal networks do not benefit from phonons of either kind as $N$ increases beyond 6. We show that, for such networks, Peierls phonons distort the optimal configurations and change the mechanism of energy transfer but do not affect the lower limit of the transfer time.

\newpage

\section{Calculation details}

\subsection{Network properties}

We consider a system of $N$ qubits, including an input site and an output site placed on the opposite sides of a cube with edge length $D$, with the remaining qubits arranged 
within the cube. Each qubit represents a two-level system with excitation energy $\epsilon_i$.  We denote the position of qubit $i$ by $\bm r_i$ and the coupling between sites $i$ and $j$ by $V_{ij}(\bm r_i, \bm r_j)$. Each qubit is coupled to local Holstein and nonlocal Peierls phonons. 
The full Hamiltonian $H=H_S+H_{B} +H_{SB}$ includes the following terms:
\begin{align}
         H_S &= \sum_{i=1}^N \epsilon_i a^{\dagger}_i a_i + \sum_{\substack{i=1,\\j>i}}^N V_{ij} (a^{\dagger}_i a_j + a^{\dagger}_j a_i),\\
         H_B &= \sum_{i=1}^{N}\hbar\omega_{i,H}b_{i,H}^\dagger b_{i,H} + \hbar\omega_{i,P}b_{i,P}^\dagger b_{i,P},\\
        \begin{split}
        H_{SB} &= \sum_{i=1}^{N} g_H a^{\dagger}_i a_i(b_{i,H}^\dagger + b_{i,H})\\
        &+ \sum_{\substack{i=1 \\ j>i}}^{N} g_P(a^{\dagger}_ia_j + a^{\dagger}_ja_i )(b_{i,P}^\dagger + b_{i,P} - b_{j,P}^\dagger - b_{j,P}), \label{eq: Peierls}
        \end{split}
\end{align}
where $a_i^{\dagger}$  and $b_i^{\dagger}$ create an excitation and phonon, respectively, on site $i$. 
We restrict the Hilbert space for the central system to a single excitation and work in the site basis $\{ \ket{i}, i=1\dots N\}$ which indicates that the excitation is on site $i$. For Hamiltonian $H_S$, we assume all site energies are the same and set $\epsilon_i=0$. The $V_{ij}$ couplings are assumed to be of the form 
\begin{equation}\label{eq:V_ij}
    V_{ij}=\frac{c}{r_{ij}^3},
\end{equation}
where $c$ is a real constant, and $r_{ij}=|\bm{r}_{j} - \bm{r}_{i}|$ are the distances between sites $i$ and $j$. 
The coupling strengths are chosen to simulate the dipole-dipole interactions responsible for exciton transport in Frenkel systems. 
As we treat both $\bm r_i$ and $\bm r_j$ as variables, we omit the anisotropy of the dipole-dipole interaction without loss of generality.

For $H_B$ and $H_{SB}$, we consider a single vibrational mode with three phonon states on site $i$ with  frequency $\omega_{i,H}$ for Holstein phonons or $\omega_{i,P}$ for Peierls phonons. The first term $b_{i,H}^\dagger + b_{i,H}$ in Eq. (\ref{eq: Peierls}) is the dimensionless displacement of a Holstein phonon coupled to site $i$. The second term $b_{i,P}^\dagger +b_{i,P} - b_{j,P}^\dagger - b_{j,P}$ is the relative displacement between sites $i$ and $j$. The Peierls phonons directly modulate the hopping amplitude $V_{ij}$ between each pair of sites. For given phonon frequencies, the coupling parameters $g_H$ and $g_P$ determine the  interactions with Holstein and Peierls phonons, respectively. The coupling parameter $g_P$ in Eq. (\ref{eq: Peierls}) is chosen as $g_P=\alpha/r^3_{ij}$, where $\alpha$ is a fixed parameter. We scale $g_P$ by $1/r_{ij}^3$ to account for both nearest-neighbour and long-range interactions with phonons. This is different from the conventional Peierls model in one dimension that allows only nearest neighbour couplings. 
To reduce the dimensionality of the problem, we restrict the motion of sites giving rise to Peierls phonons along the dimension joining the input and output sites.

The effect of coupling to phonons in lattice systems can be quantified by the dimensionless parameter $\lambda = 2g^2/(\hbar \omega_{\rm ph} t)$, where $g$ is the particle-phonon coupling, $t$ is the nearest neighbour hopping amplitude, and $\omega_{\rm ph}$ is the phonon frequency  \cite{lambda-couplings, sharp_trans, polaron_trans}. The values of $\lambda \gtrsim 1$ correspond to the strong coupling regime. It is, therefore, sufficient to vary either the coupling $g$ or the phonon frequency in our analysis. We choose to vary $\omega_i$ and  define an effective site-dependent coupling strength $\lambda_i = 2g_H^2/(\hbar \omega_i V_\text{max})$ for the Holstein couplings and $\lambda_i = \frac{1}{N-1} \sum_{j \neq i}^N 2 \alpha^2/(\hbar \omega_i r^6_{ij} V_\text{max})$ for the Peierls couplings, 
where $V_\text{max}$ is the maximum value of the hopping amplitude $V_{ij}$ in the disordered arrays considered here. 
The values of $\lambda_i$ thus defined are used to quantify the effective strength of phonon-induced couplings. Note that for the Peierls case, $\lambda_i$ depends on both $\omega_i$ and $r_{ij}$ that are varied independently. 
We attempted to optimize simultaneously $g_H \in [1,20]~\hbar~{\rm ps}^{-1}$, $\bm r_i$ and $\omega_i$ for the Holstein case and  $\alpha \in [1,20]~\hbar~$\AA$^3$ ps$^{-1}$, $\bm r_i$ and $\omega_i$ for the Peierls case. 
The optimal solutions converge to values of $g_H \in [1, 5]~\hbar$ ps$^{-1}$ and $\alpha \approx 15~\hbar$ \AA$^3$ ps$^{-1}$. Guided by these results, we fix $g_H = 1.45~\hbar$ ps$^{-1}$ and $\alpha = 15~\hbar$ \AA$^3$ ps$^{-1}$ for the final calculations.

\subsection{Transfer time}

We denote the input site as $|i=1\rangle = |\rm in \rangle$ and the output site as $|i=N\rangle = |\rm out \rangle$. 
It is convenient to define the transfer time in terms of half the period of the Rabi oscillation between the input and output sites in the absence of any other qubits: 
\begin{equation*}
    T = \frac{\pi}{2V_\text{min}}
\end{equation*}
where $V_{\rm min}$ is the strength of the coupling between $| \rm in \rangle$ and $|\rm out \rangle$. 
To make the excitation transfer unidirectional, we follow Ref. \cite{non-hermit} to introduce 
a non-Hermitian term $H_{\Gamma}=-i\frac{\Gamma}{2}\ket{\text{out}}\bra{\text{out}}$ with sink rate $\Gamma$ to the total Hamiltonian $H$. 
This allows one to define  the $| \rm in \rangle \rightarrow | \rm out \rangle$ transfer time as
\begin{equation}
   \mathfrak{T} =\int_0^\infty (1-\mathfrak{p}(t)) dt= \int_0^\infty \sum_i^N \rho_{ii}(t) dt,
\end{equation}
where $\mathfrak{p}(t)$ is the sink population and $\rho(t)$ is the reduced density matrix for the central system. In order to compute the time evolution of the reduced density matrix, 
\begin{align}
     \rho(t) &= \text{Tr}_B \{\ket{\Psi(t)}\bra{\Psi(t)}\}
\end{align}
where $\text{Tr}_B$ is the trace over the phonon degrees of freedom,
we diagonalize the total Hamiltonian $H$ and compute the time evolution of the full wave function as 
\begin{align}
    \ket{\Psi(t)} &= e^{-iHt}\ket{\text{in}},
\end{align}
 where $\ket{\text{in}}=\ket{e_1}\otimes \ket{g_2}\otimes\dots \otimes\ket{g_N}$ with $\ket{g_i}$ ($\ket{e_i}$) being the ground (excited) state at site $i$.

To put our model in the context of the FMO system, we scale the parameters by real physical units. The input and output sites are $D=200$ \AA\ apart, which is roughly the diameter of the monomer. The minimal distance between neighbouring intermediate sites is 5 \AA\, accounting for the size of each complex. The excitonic coupling constant $c$ in Eq. (\ref{eq:V_ij}) is set to $c=10^{-6}~\hbar$ \AA$^3$ ps$^{-1}$, and the corresponding 2-site benchmark time scale is $T=12.5$ ps. The sink rate $\Gamma$ is expressed in units of $1/T$. All phonon frequencies have the unit of ps$^{-1}$. In the following discussion, the transfer time and any time scales are written as a fraction of $T$.

\subsection{Bayesian optimization}

The transfer time $\mathfrak{T}$ described above depends on the configurations of intermediate sites $\bm R = [\bm r_2, \cdots, \bm r_{N-1}]^\top$, phonon frequencies $\bm \omega_H = [\omega_{1,H}, \cdots, \omega_{N,H}]^\top$ and $\bm \omega_P = [\omega_{1,P}, \cdots, \omega_{N,P}]^\top$, and the corresponding exciton-phonon coupling strengths. Our goal is to find the specific combinations of these parameters that minimize $\mathfrak{T}$. We adapt Bayesian optimization (BO), a global non-convex black-box function optimization method, to explore this parameter space. The site coordinates are sampled from $x_i,y_i,z_i \in [-\frac{D}{2},\frac{D}{2}]$, with the constraint  $|\bm r_i-\bm r_j|>r_\text{min}$.
Since we vary the phonon frequencies on each site independently, only one value of $g$ is chosen for all sites. The phonon frequencies are sampled from $\omega_i \in [1.25, 125] \,{\rm ps}^{-1}$. The sink rate $\Gamma$ is set as a predetermined value for each optimization run as described in the following section. 

Bayesian optimization is a machine learning algorithm for optimizing functions that do not have a closed-form expression \cite{BO2}. It is particularly useful when the objective function is expensive to evaluate. For a discussion of applications of BO to quantum problems, see Ref. \cite{bo-pccp}.
The main idea behind BO is to build a probabilistic model for the target function $f(\bm{\theta})$, where $\bm{\theta}$ includes all the variable parameters. The optimization is initiated with a set of (randomly selected) pairs $\{\bm{\theta}_i,\mathfrak{T}_i\}$ by constructing a probabilistic model over this data set. The algorithm then proposes the next evaluation at $\bm \theta_\text{new}$ based on an acquisition function (see below). The algorithm updates the function values iteratively until the convergence criteria are met. One of the most important steps in BO is to construct the probabilistic model for function $f$. In the present work, this is done by modelling  $f$ with a Gaussian Process (GP) \cite{rasmussen-book, bo-pccp}. 
GP is a generalization of Gaussian distribution over random variables to distribution over functions, specified by the mean and covariance functions. 

Training a GP amounts to conditioning the mean and covariance functions by observed data. 
In the present work, we parametrize the GP covariance as 
\begin{equation*}
    k(\bm{\theta}, \bm{\theta}') = \left[1+\frac{d(\bm{\theta},\bm{\theta}')^2}{2\gamma l^2}\right]^{-\gamma},
\end{equation*}
and determine the hyperparameters $\gamma$ and $l$ by maximizing the logarithm of the marginal likelihood of the model \cite{rasmussen-book, bo-pccp}. 

The acquisition function ${\cal A}(\bm{\bm{\theta}})$ used for BO in the present work is defined by the conditional mean $\mu$ and variance $\sigma^2$ of the trained GP as
\begin{equation}
    {\cal A}(\bm{\bm{\theta}}) = \mu(\bm{\bm{\theta}}) - \kappa \sigma(\bm{\bm{\theta}}),
\end{equation}
where $\kappa$ is a fixed parameter for exploration-exploitation control. The black-box function $f$ is evaluated at each iteration at $\bm{\theta}_\text{new}$ that maximizes $\cal A$. Note that the acquisition function takes into account both the predicted value $\mu$ and the uncertainty $\sigma$ to propose the next value. The total number of function evaluations $f$ depends on the dimensionality of the problem but can be estimated as $\approx 10 \times {\cal D}$, where ${\cal D}$ is the number of independent variables. 
For examples of recent applications of BO to quantum problems, illustrating in particular the speed of convergence, see Refs. \cite{bo-1,bo-2}.

\section{Results}







\subsection{Phonon-free networks}
\begin{figure}[t]
    \centering
    \includegraphics[width=10cm]{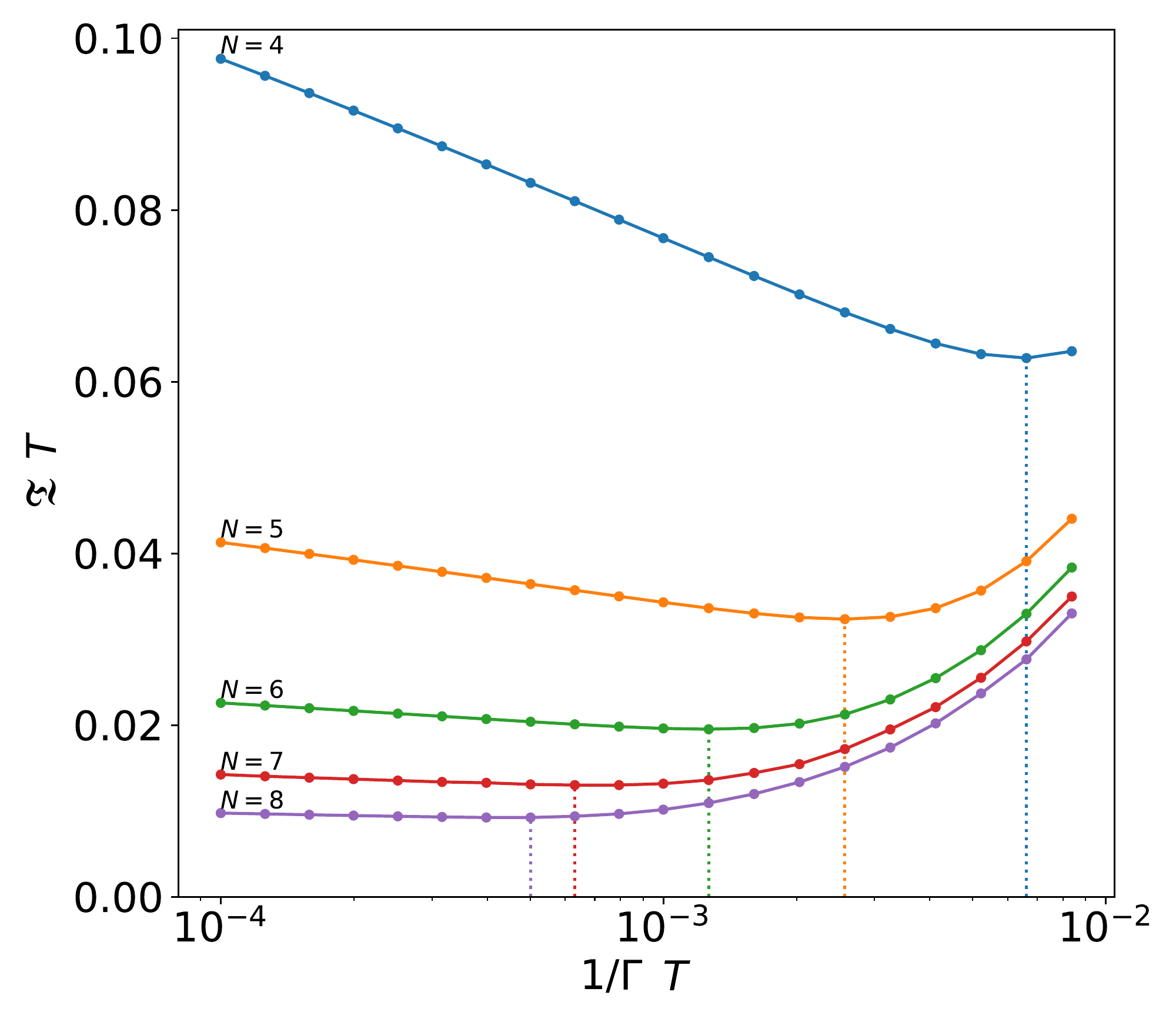}
    \caption{Minimal transfer time $\mathfrak{T}$ as a function of inverse sink rate $\Gamma$ without coupling to phonons for different numbers of sites  ($N=4-8$).  The global optimal sink rates and the corresponding transfer times are marked by vertical dotted lines.}
    \label{fig:Trans_T_Ns}
\end{figure}

 We first study the optimal networks without phonon couplings in order to compare the results of BO with the previous work \cite{optnet1,optnet2}. As illustrated previously by Scholak \textit{et al.} \cite{optnet1}, the minimal transfer time depends both on the number of intermediate sites and the non-Hermitian sink rate at the boundary. Figure \ref{fig:Trans_T_Ns} shows the minimal transfer time $\mathfrak{T}$ as a function of $1/\Gamma$ for several $N$-site configurations. Each point of each curve is obtained by minimizing $\mathfrak{T}$ using BO. Multiple runs of BO repeated with different initial conditions converge to the same values. In agreement with Ref. \cite{optnet1}, Figure \ref{fig:Trans_T_Ns} shows that the transfer time decreases as $N$ increases at a given sink rate $\Gamma$. This $N$ dependence is prominent at a large sink rate, and the minimal transfer time eventually converges to the $N=2$ limit at low values of $\Gamma$, as shown by Scholak \textit{et al.} \cite{optnet1}. In addition, $\mathfrak{T}$ first decreases to a minimum value and then increases with an increase of $\Gamma$, which is the manifestation of quantum Zeno effect \cite{zeno}. Hence, there exists a global optimal $\Gamma$ for each $N$ that minimizes the transfer time for all possible configurations. Unless specified otherwise, we fix the value of $\Gamma$ to correspond to the minimum transfer time through the phonon-free network with the corresponding number of sites, i.e. the values that minimize the curves in Figure \ref{fig:Trans_T_Ns}, shown  in Figure \ref{fig:Trans_T_Ns} by dotted vertical lines.  
 
\begin{figure}[h!]
    \centering
    {{\includegraphics[width=16cm]{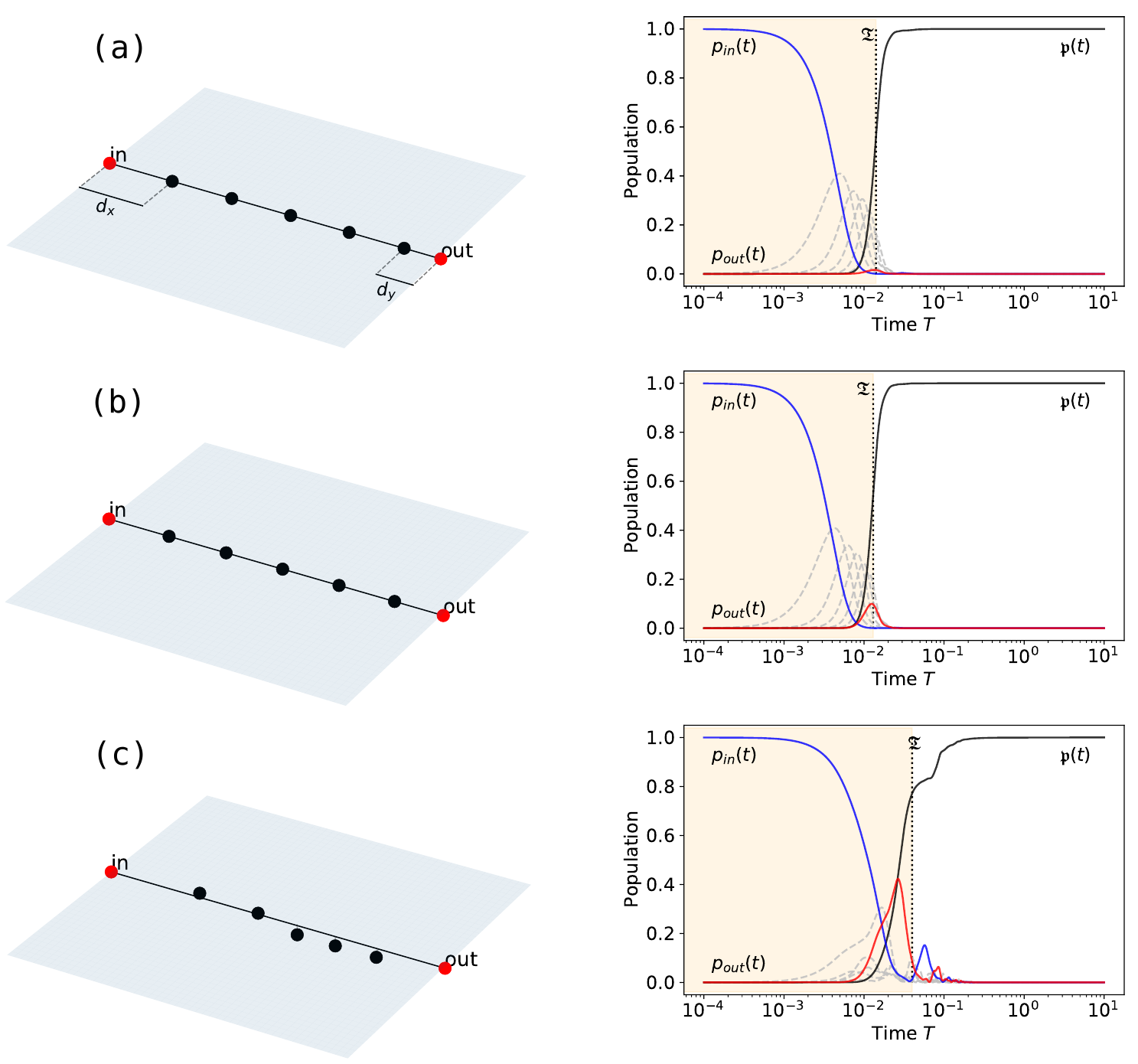} }}
    \caption{Left panels: Optimal configurations at (a) $1/\Gamma=10^{-4}~T$, (b) $1/\Gamma=6.3 \times 10^{-4}~T$ and (c) $1/\Gamma=10^{-2}~T$. Note the distortion $d_x>d_y$ in (a).
    Right panels: the population dynamics for the input-site population $p_{in}(t)$ (blue solid line), output-site population $p_{out}(t)$ (red solid line) and intermediate sites (grey dashed line). The solid black line represents sink population $\mathfrak{p}(t)$. The vertical dotted line shows transfer time $\mathfrak{T}$, and the shaded area is the time range within $\mathfrak{T}$.
        }
    \label{fig:7s_NCoh_config}
\end{figure}

As illustrated by Scholak \textit{et al.} \cite{optnet1}, the configurations corresponding to the minimal transfer time at large $\Gamma$ tend to be linear. 
Figure \ref{fig:7s_NCoh_config} shows three examples of $N=7$ configurations produced by BO for different values of $\Gamma$, with (b) illustrating the configuration corresponding to the global minimum in transfer time.  As expected, the optimal configuration is both linear and periodic, leading to the analogue of ballistic transport, as illustrated in (b), right.   By comparison with (b), the configuration in (a) is slightly shifted towards the output site, while the configuration in (c) is disordered. Both distortions perturb ballistic transport and decrease the transfer time, as manifested by the more complex dynamics of different site populations (depicted in the right panels).  


\subsection{Networks coupled to Holstein phonons}
\begin{figure}[p!]
    \centering
    {{\includegraphics[width=16cm]{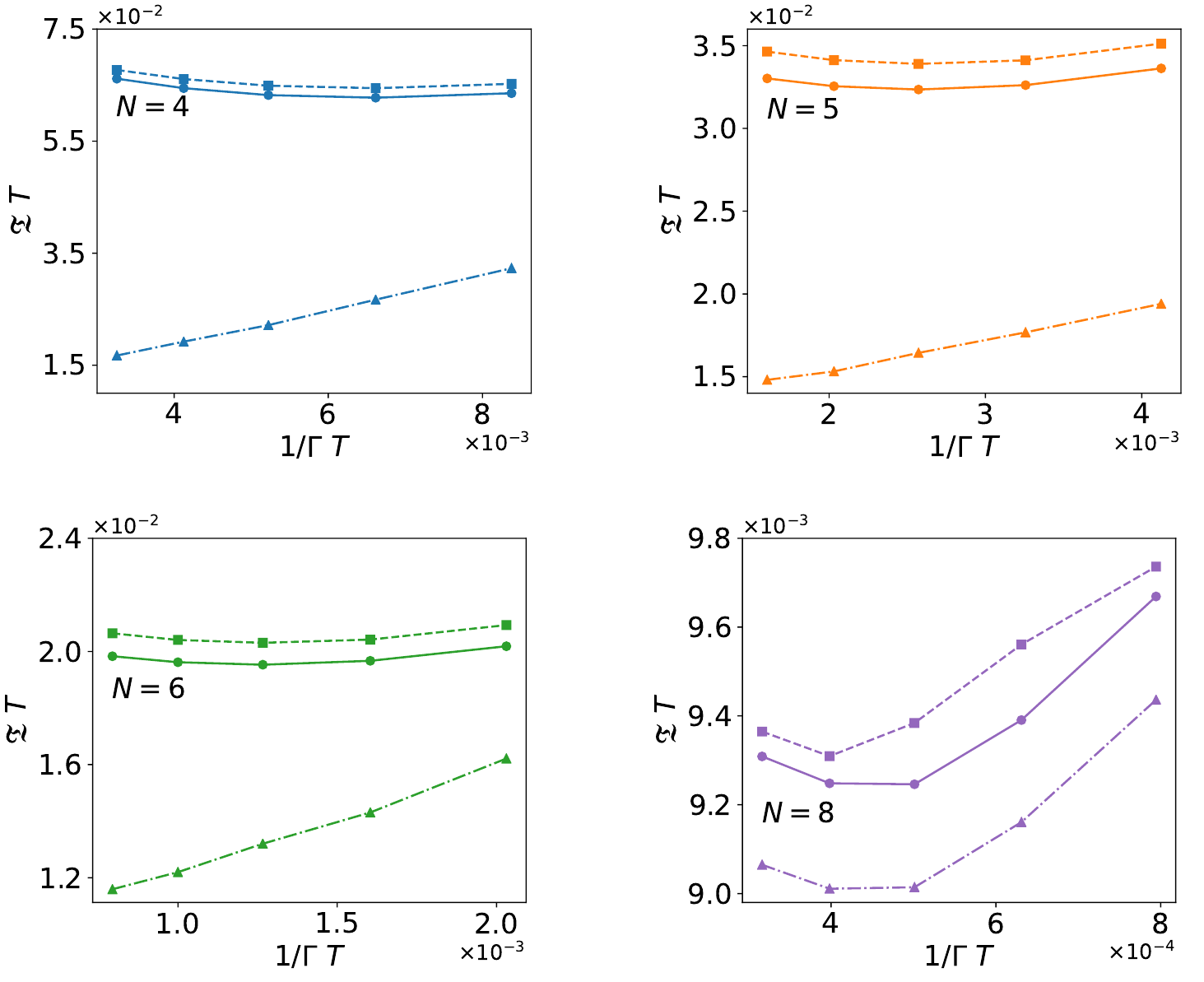} }}
    \caption{Minimal transfer time $\mathfrak{T}$ as a function of the inverse sink rate $1/\Gamma$ for networks with different numbers of sites ($N=4, 5, 6, 8$) in the vicinity of the globally optimal $\Gamma$ as defined by the bare phonon cases in Figure \ref{fig:Trans_T_Ns}: circles --  no couplings to phonons; squares -- minimal transfer times with Holstein couplings; triangles -- minimal transfer times with Peierls couplings.}
    \label{fig:Trans_T_Ns_Open_cases}
\end{figure}

In this section, we couple each qubit to local Holstein phonons and optimize simultaneously the site positions and each of the site phonon frequencies. Figure \ref{fig:Trans_T_Ns_Open_cases}  compares the $\Gamma$-dependence of the minimal transfer time for networks with and without phonons. The squares are minimal transfer times under Holstein coupling in the vicinity of the globally optimal sink rate. The transfer times follow the trend observed in the phonon-free case but exhibit larger magnitudes. The optimization algorithm tends to decouple the phonons and converges to the same site configurations as in the phonon-free case. This illustrates that Holstein couplings cannot produce networks with accelerated transfer time, if network topology is allowed to be optimized.

\begin{figure}[p!]
    \centering
    {{\includegraphics[width=10cm]{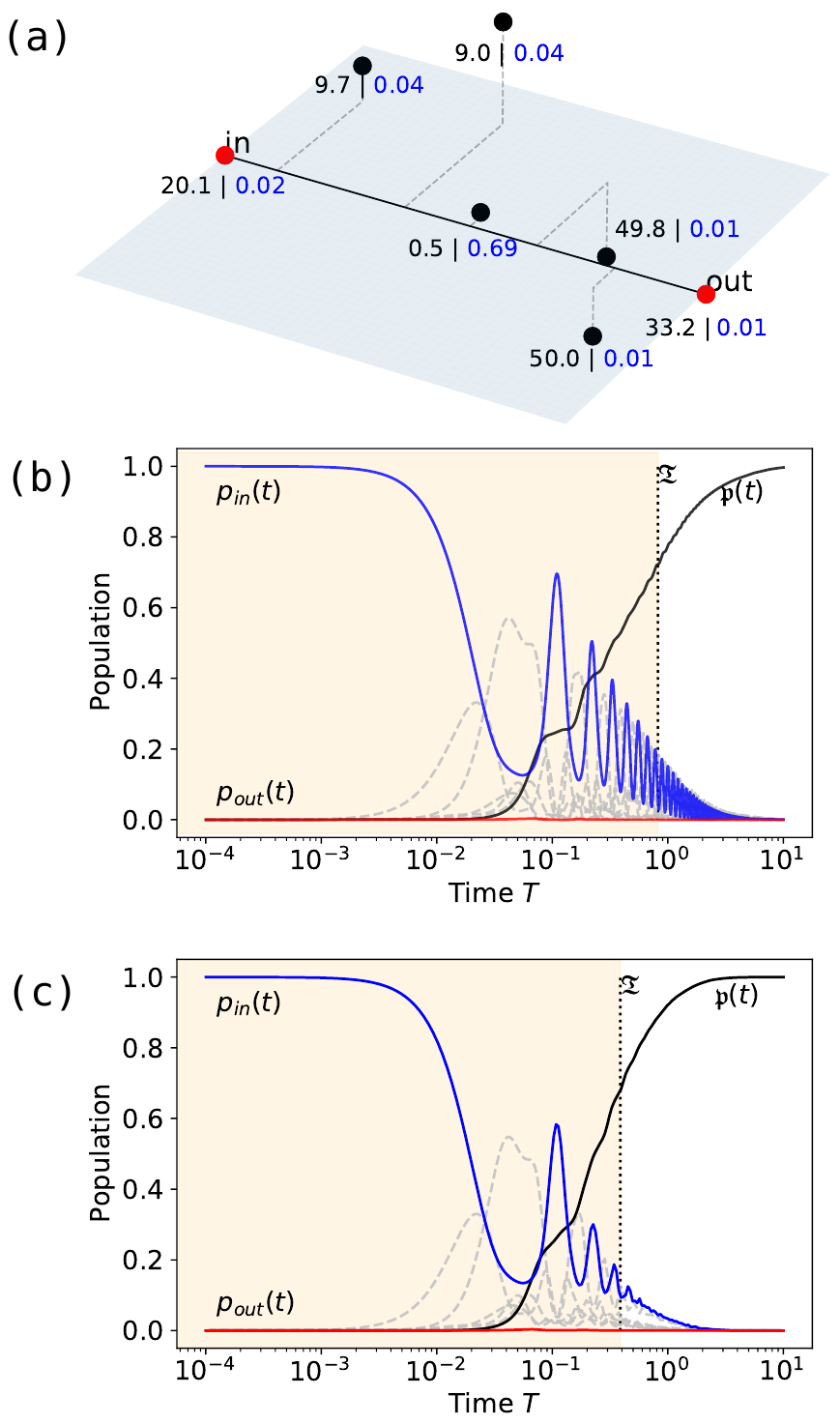}}}
    \caption{(a) A random 3D configuration with $1/\Gamma = 6.3\times 10^{-4}~T$ coupled to Holstein phonons.
    Each site is labeled by $(\omega_i\,|\, \lambda_i)$, where $\omega_i$ are
the optimal Holstein frequencies (in ${\rm ps}^{-1}$) determined by BO, and $\lambda_i$ (shown in blue) are the corresponding dimensionless
    effective coupling strengths.
    (b) Population dynamics without Holstein couplings: $\mathfrak{T}=0.688~T$. 
    (c) Population dynamics with Holstein coupling $g_H = 1.45~\hbar~{\rm ps}^{-1}$ to phonons with frequencies shown in (a):  $\mathfrak{T}=0.365~T$.
    The blue, red (solid) and grey (dashed) lines are the populations at the input, output and intermediate sites.
    }
    \label{fig:7s_holstein}
\end{figure}

As discussed in the introduction, many theoretical studies based on quantum master equation approaches found that Holstein couplings can accelerate the excitation energy transport compared with the purely coherent cases \cite{EAQT3,efftrans,fmorev}. These studies always consider a fixed system Hamiltonian coupled with a Holstein bath. We can explore a similar scenario by coupling a particular network with fixed site positions to phonons with variable, site-dependent frequencies. Figure \ref{fig:7s_holstein} shows a randomly selected 3D network of $N=7$ sites. We observe that optimizing the phonon frequencies of this network with fixed site positions reduces the transfer time in half. 

As seen in the population dynamics plots of Figure \ref{fig:7s_holstein}, in the absence of phonons, the populations oscillate between the input and intermediate sites with high frequency due to localization near the input site. The Holstein couplings accelerate the transfer by reducing the interference and smoothing out the high-frequency component. Such oscillations are absent in optimal phonon-free geometries. As a result, adding Holstein couplings leads to suppression rather than enhancement of transfer time for networks with optimal phonon-free configurations.



\subsection{Networks with Peierls coupling}
\begin{figure}[t]
    \centering
    \includegraphics[width=10cm]{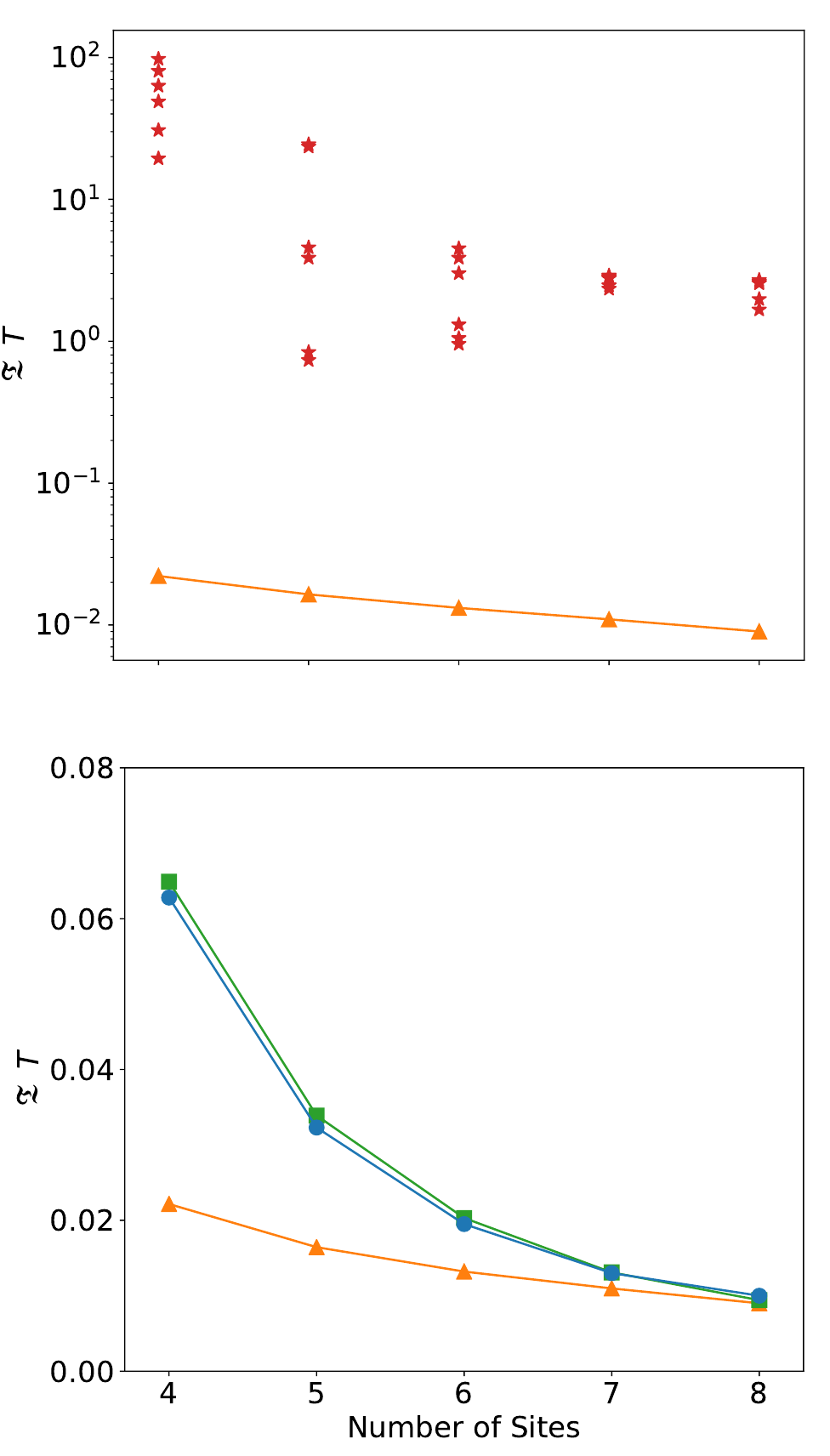}
    \caption{Lowest transfer time vs the number of sites for optimal networks: circles -- phonon-free, squares -- with Holstein phonons, triangles -- with Peierls phonons,  stars -- classical transport. The stars are transfer time for optimal non-linear Peierls configurations, and only the fastest 6 paths are shown in the plot.}
    \label{fig:Non_SSH_Hol_Num_n}
\end{figure}

\begin{figure}[p!]
    \centering
    {{\includegraphics[width=16cm]{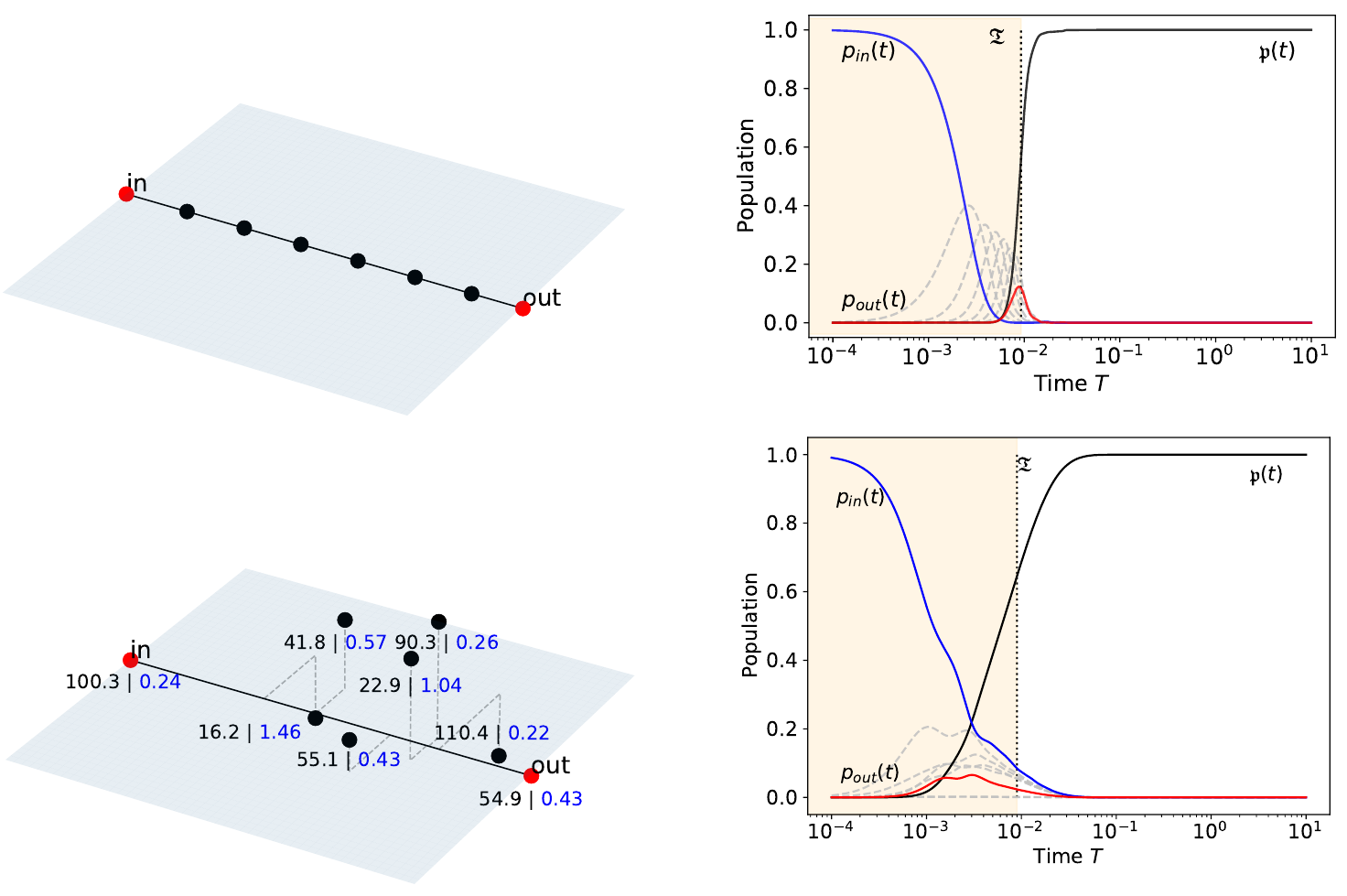} }}
    \caption{Comparison between the optimal phonon-free network (top panels) and optimal network with $\alpha=15$ $~\hbar$ \AA$^3$ ps$^{-1}$ (bottom panels). 
    The two configurations produce a similar optimal transfer time $\mathfrak{T}=0.009~T$ at $1/\Gamma=5.0 \times 10^{-4}~T$. 
    Each site is labeled by $(\omega_i\,|\, \lambda_i)$, where $\omega_i$ are
the optimal Peierls frequencies (in ${\rm ps}^{-1}$) determined by BO, and $\lambda_i$ (shown in blue) are the corresponding dimensionless
    effective coupling strengths.
    } 
    \label{fig:SSH_8s_config}
\end{figure}

Unlike Holstein couplings, Peierls couplings lead to significant changes in population dynamics and optimal configurations. Triangles in Figure \ref{fig:Non_SSH_Hol_Num_n} show the minimal transfer time through networks with Peierls couplings. Note that, for $N=4-6$, the globally minimal $\mathfrak{T}$ no longer corresponds to the same optimal value of  $\Gamma$ as in  the phonon-free cases. However, we continue to use the optimal values of $\Gamma$ from Figure \ref{fig:Trans_T_Ns} for consistency.
Figure \ref{fig:Non_SSH_Hol_Num_n} shows that, for $N=4-6$, there is a clear decrease in transfer time due to Peierls couplings, whereas, for $N=7-8$, the values of $\mathfrak{T}$ with and without Peierls phonons are the same (see also Figure \ref{fig:SSH_8s_config}). 
Note that the optimal network configurations leading to the same transfer time with and without Peierls couplings for $N > 6$ are completely different, as illustrated in Figure \ref{fig:SSH_8s_config} for the $N=8$ case.

\begin{figure}[p!]
    \centering
    {{\includegraphics[width=16cm]{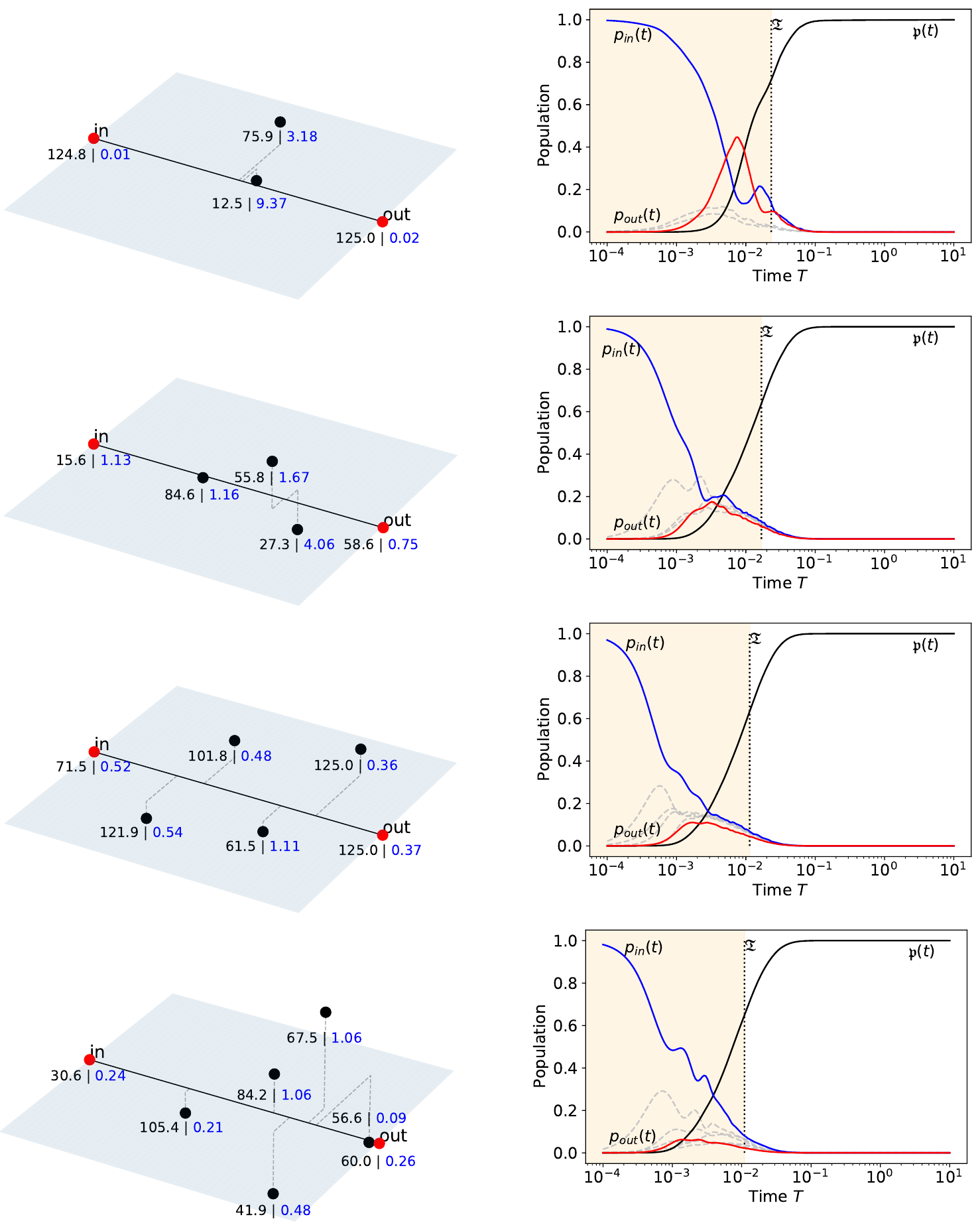} }}
    \caption{Optimal configurations (left) and corresponding population dynamics (right) for networks with $N=4-7$ coupled to Peierls phonons.
    The coupling constant $\alpha=15$ $~\hbar$ \AA$^3$ ps$^{-1}$. 
    Each site is labeled by $(\omega_i\,|\, \lambda_i)$, where $\omega_i$ are
the optimal Peierls frequencies (in ${\rm ps}^{-1}$) determined by BO, and $\lambda_i$ (shown in blue) are the corresponding dimensionless
    effective coupling strengths.
    } 
    \label{fig:SSH_config}
\end{figure}

The phonon-free configurations realize the minimal transfer time by linear or quasi-linear arrangements shown in Figures \ref{fig:7s_NCoh_config} and \ref{fig:SSH_8s_config} (upper panel). With Peierls couplings, however, the optimal configurations are no longer linear, as shown in Figure \ref{fig:SSH_8s_config} (lower panel). Figure \ref{fig:SSH_config} shows several examples of optimal networks with Peierls couplings. Starting from $N=4$, Peierls couplings force the intermediate sites to be placed into 3D arrangements between the input and output sites. There are several other network configurations that exhibit similar transport efficiency. Some of these configurations are related by rotational symmetry, while others are not physical due to the clustering of sites. 
The optimal Peierls phonon frequencies are not unique either. Phonon frequencies labelled in Figure \ref{fig:SSH_config} exhibit no particular patterns, with most phonon frequencies in the range between $50-100~{\rm ps}^{-1}$. The values of  $\lambda_i$ are generally significant, indicating strong coupling to Peierls phonons. It is thus clear that lattice site vibrations analogous to those causing Peierls distortion \cite{Peierls_dis} play a critical role in reducing the transfer time through non-linear networks.

Another feature of networks with Peierls couplings can be observed from the population dynamics. In the large $\Gamma$ limit, we see ballistic-like transport for the phonon-free lattice with 
the excitation populating each site in sequence. 
 In Figure \ref{fig:SSH_config}, however, all intermediate sites are populated in parallel  within a narrow time interval. The population of the input site decreases monotonously. For some of the optimal networks (e.g. $N=4, 6$), we observe a strong overlap of the population dynamics of the intermediate sites, indicating that the excitation becomes maximally delocalized. 
This shows that the optimal configurations of such networks ensure that Peierls couplings collectively enhance the effect of the hopping amplitudes between all pairs of sites. In general, Peierls couplings are known to modulate the kinetic energy of electrons, resulting in light Su-Schrieffer-Heeger polarons  \cite{sharp_trans, light_pol_1, light_pol_2} and bipolarons \cite{lambda-couplings} at strong electron-phonon coupling, by bringing sites closer together to allow for enhancement of particle hopping.  The results in Figure \ref{fig:SSH_config} are the manifestation of the same phenomenon.
 

The above results are obtained with three phonon states per site. Some of the resulting configurations have $\lambda_i>1$, which indicates that more phonon states may be necessary for convergence. To explore the effect of this phonon state truncation, we repeated the calculations for 10 optimized networks with $N=7$ sites and four phonons per site. 
This expansion of the phonon state space, while increasing the truncated space from 3 states to 4 states, changes the transfer time by less than 15 \% and does not result in any qualitative changes. In particular, the general trend illustrated in Figure \ref{fig:Non_SSH_Hol_Num_n} (lower panel) remains the same with 4 phonon states per site.


\subsection{Classical vs quantum transport}

It is instructive to compare the optimal transfer times obtained by BO in the previous subsections with the time of classical energy transfer. 
We consider F\"{o}rster resonance energy transfer theory as the classical transport model \cite{Forster1, Forster2} with the electronic coupling between a donor and an acceptor 
 chosen as $V_{ij}$.
As the excitation transfer involves two sites at a time, the input and output sites of a network can be viewed as connected by a manifold of possible classical paths. 
 We assume that the excitation traverses all sites in the configurations optimal with Peierls couplings and present six shortest transfer times.

 Figure \ref{fig:Non_SSH_Hol_Num_n} (upper panel) plots the classical results for a comparison with the quantum transfer time in optimal networks with Peierls couplings. The fastest 6  classical transfer times are more than two orders of magnitude longer than the quantum transfer times for $N=7,8$. The shortest classical transfer time of the linear configuration is the same as the quantum transfer. This is expected as Figure \ref{fig:7s_NCoh_config} shows that quantum transport is ballistic in this case. The minimal classical transfer time through the optimal Peierls configurations, however, is two orders of magnitudes longer than the corresponding quantum transfer times. 

\section{Summary}
In this work, we explore the efficiency of quantum energy transfer through finite networks of qubits coupled to phonon ensembles. 
We consider a microscopic model of both the networks and phonons, treating the position of each individual qubit and each site-dependent phonon frequency as independent variables. 
The transfer time is computed by rigorous quantum dynamics calculations 
and optimized by simultaneously varying these model parameters. We consider two types of phonons: Holstein phonons that model an external bosonic field and Peierls phonons that modulate the distance between qubit sites in the networks. 
Our calculations allow us to consider networks with up to $N=8$ qubits, each coupled to three vibrational states. 
We observe the following interesting results:

\begin{itemize}

\item If the qubit positions are fixed in a random, suboptimal configuration, Holstetin couplings may decrease the transfer time. 

\item If both the qubit positions and the Holstein frequencies are optimized simultaneously, the optimal solution giving the lowest transfer time yields networks decoupled from phonons, for any value of $N$. We thus conclude that Holstein couplings cannot accelerate quantum transport beyond the limit of the transfer time in an optimal phonon-free configuration. 

\item If both the qubit positions and the Peierls frequencies are optimized simultaneously, the optimal solutions for $N=2-6$ yield configurations distorted by Peierls couplings to produce faster transfer times than through phonon-free networks with the same number of sites $N$. 

\item The acceleration of transfer time by Peierls couplings decreases with $N$ and disappears for $N=8$. We thus conclude that, while Peierls couplings may accelerate transfer through very small networks, they offer no advantage for networks with $N>7$. It is important to note that the Peierls couplings change the optimal network configurations, even for $N>7$. This is clearly illustrated by Figure 6 presenting two optimal networks of $N=8$ qubits with (lower panels) and without (upper panels) Peierls couplings. While the transfer time through these networks is the same and corresponds to the minimum possible time, quantum dynamics is very different, changing from ballistic (in the upper panels) to one that populates multiple sites simultaneously (in the lower panels). 

\item All of the optimal quantum transfer times are significantly faster than the corresponding classical times, as calculated using F\"{o}rster theory, for non-linear network geometries.

\end{itemize}

Beyond the above results, the present work illustrates an application of Bayesian optimization for the design of quantum networks and the exploration of quantum speed limits. 
The present approach can be extended to determining the realistic values of the quantum evolution speed for applications, where theoretical bounds overestimate the limits \cite{quantum-speed-limit}. 
The present approach can also be extended to optimization of quantum transfer through systems coupled to large baths. Instead of optimizing the individual phonon frequencies as in the present work, one would optimize the parameters of the bath spectral functions while solving the master equation for the reduced density matrix.  In the case of unknown spectral functions, one can model them with neural networks and apply reinforcement learning strategies.

\section*{Data availability}

The data that support the findings of this study are available from the corresponding author upon reasonable request.

\section*{Acknowledgments}
This work was supported by NSERC of Canada.

\clearpage
\newpage

\end{document}